\documentstyle[twocolumn,pre,aps]{revtex}
\input{psfig.sty}
\input epsf.sty
\begin{document}
\draft
\title{Bulk and Interfacial Shear Thinning of Immiscible Polymers}
\author{Sandra Barsky }
\address{ Lawrence Berkeley National Laboratory and Dept. of Mathematics,
 University of California at Berkeley, Berkeley, CA 94720}
\author{Mark O. Robbins}
\address{Department of Physics and Astronomy, Johns Hopkins University,
Baltimore, MD 21218}

\date{\today }
\maketitle

\begin{abstract}
Nonequilibrium molecular dynamics simulations are used to 
study the shear thinning behavior of immiscible symmetric polymer blends.
The phase separated polymers are subjected to a simple shear flow imposed by
moving a wall parallel to the fluid-fluid interface.
The viscosity begins to shear thin at much lower rates in the bulk
than at the interface.
The entire shear rate dependence of the interfacial viscosity is
consistent with a shorter effective chain length $s^*$ that
also describes the width of the interface.
This $s^*$ is independent of chain length $N$ and is a function only of
the degree of immiscibility of the two polymers.
Changes in polymer conformation are studied as a function of position
and shear rate.
Shear thinning correlates more closely with a decrease in
the component of the radius of gyration along the velocity gradient
than with elongation along the flow.
At the interface, this contraction of chains is independent of $N$
and consistent with the bulk behavior for chains of length $s^*$.
The distribution of conformational changes along chains is also studied.
Central regions begin to stretch at a shear rate that decreases
with increasing $N$, while 
shear induced changes at the ends of chains are independent of $N$.
\end{abstract}

\section{Introduction}\label{sec:intro}

The viscosity of a polymer melt decreases when the melt is subjected to a 
sufficiently large shear rate, a phenomenon known as shear thinning.
This non-Newtonian behavior is important in polymer processing and
applications and has been extensively studied both experimentally and
theoretically\cite{carl,ganaz,semenov}.
The basic origin of shear thinning is that polymers elongate
and align with the flow when they are sheared
more rapidly than they can relax.
As the shear rate increases, the degree of alignment rises and
the polymers present a decreasing hydrodynamic resistance to flow.
Recent numerical simulations \cite{liu,lopez,daivis,xu,lyulin}
have investigated a variety of issues in this non-Newtonian regime,
including the shear-rate dependence of the viscosity, polymer 
conformation, and wall slip.

One area which has yet to be explored is shear-thinning near the
interface between immiscible polymers.
Blending of polymers is common in industry, as this is 
one way of creating new materials.
However, the blended polymers are often immiscible, leading to phase separation.
The resulting interfaces can have a major impact on the processing
and ultimate properties of the blend.

The static properties of interfaces in binary polymer blends are
typically described by the Flory-Huggins model \cite{flory}.
The degree of immiscibility is characterized by a parameter $\chi$,
which represents the free energy cost for placing a monomer of one
type into a homogeneous region of the other type.
There is a characteristic length $s^*$ of the loops of one type of polymer
that penetrate into the region occupied by the other polymer. 
By balancing the enthalpic cost and entropy gain, one finds that
$s^* \sim 1/\chi$ \cite{degennes}.
The interface width scales as the radius of gyration of a segment of length
$s^*$.

De Gennes and coworkers argued that the dynamic properties of the interface
should also be controlled by $s^*$ rather than the chain length $N$
\cite{degennes}.
In particular, they suggested that for polymers in the Rouse limit
the interfacial viscosity $\eta_I$
should be given by the bulk viscosity of chains of length $s^*$.
This implies that $\eta_I$ is only a function of $\chi$ and has
no dependence on $N$ (as long it is larger than $s^*$).
Goveas and Fredrickson reached similar conclusions from more detailed
calculations \cite{goveas}. 

In a recent paper we presented a simulation study of the interfacial
viscosity in the Newtonian regime of a symmetric
binary blend of Rouse polymers \cite{barsky}.
As predicted \cite{degennes,goveas}, we
found that the interfacial viscosity is determined by
the degree of immiscibility and converges to a chain length independent
value at large $N$.
Moreover, both $\eta_I$ and the interface width decrease with increasing
immiscibility and their values are consistent with a common
value of $s^*$.

In this paper we investigate the shear thinning behavior of the interfacial 
viscosity.
We find that the same $s^*$ that sets the width of the interface
and the Newtonian viscosity determines the shear rate where shear
thinning begins.
Indeed the entire variation of $\eta_I$ with shear rate follows the
bulk shear thinning curve for chains of length $s^*$ and is independent of $N$.
The shear rate dependent viscosity converges to the bulk behavior for
chains of length $N$ at distances of order $R_g$ from the interface.
Since $N$ is larger than $s^*$,
the bulk viscosity $\eta_B$ is higher than $\eta_I$ in the Newtonian
regime and begins to shear thin at a lower shear rate.
Indeed, the interface does not begin to shear thin until $\eta_B$ has
decreased to a value that is comparable to the
Newtonian limit of $\eta_I$.
At higher shear rates the viscosity is nearly independent of distance
from the interface.

We also examine changes in the conformational properties of chains 
near the interface and in the bulk as a function of shear rate.
In equilibrium, the bulk values of the mean-square components of the
end-to-end vector are equal, $R^2_x = R^2_y=R^2_z$.
The two components in the plane of the interface ($x$ and $y$) remain
nearly unchanged at the interface, but the normal component decreases
because of the constraint imposed by immiscibility.
As the shear rate is increased, polymers elongate along the direction
of flow ($x$), and contract in the orthogonal directions.
We find substantial elongation before there is any measurable change in
bulk viscosity.
The onset of bulk shear thinning appears to be more closely correlated with
the onset of contraction in the $z$ direction.
The correlation between viscosity and contraction is even more dramatic
at the interface.
Like $\eta_I$, plots of $R^2_z$ as a function of shear rate
are independent of $N$ and follow the bulk behavior for chains of length $s^*$.
In contrast, changes in elongation depend on $N$ and begin at lower
shear rates than changes in $\eta_I$.

The remainder of the paper is organized as follows.  In
the next section we review the interaction potentials and simulation
techniques used in this work.
Results and analyses are described in Sec. \ref{sec:result}, and 
a summary and discussion are presented in Sec. \ref{sec:dis}.

\section{Model}\label{sec:method}

The polymer model and simulation techniques are similar to those used in
previous work \cite{barsky,robbins,thompson}.
The polymer potential is based on the bead-spring model 
developed by Kremer and Grest \cite{grest1}.
Linear polymers containing $N$ beads each are created by linking
nearest  neighbors on a chain with the potential 
\begin{equation}\label{vgrest}
U_{nn}(r_{ij})=\left\{\begin{array}{ll}-\frac{1}{2}kR_0^2\ln\left[1-
\left(r_{ij}/R_0\right)^2\right]&r_{ij}<R_0\\ \infty &
r_{ij}\geq R_0\,,
\end{array}\right.\end{equation}
where $r_{ij}$ is the distance between beads $i$ and $j$, $R_0=1.5\sigma$,
$k=30\epsilon/\sigma^2$, and $\sigma$ and $\epsilon$ set the length and energy 
scales, respectively.
All particles in the system interact through a truncated Lennard-Jones potential
\begin{equation}\label{LJ}
U_{LJ}(r_{ij})=\left\{\begin{array}{ll}4\epsilon_{\alpha\beta}\left[\left(
\sigma/r_{ij}
\right)^{12}-\left(\sigma/r_{ij}\right)^{6} \right]&r_{ij}<r^c_{\alpha\beta}\\
 0&r_{ij}\geq r^c_{\alpha\beta}\,, \end{array} \right.
\end{equation} 
where the interaction energy $\epsilon_{\alpha\beta}$ and cutoff
$r^c_{\alpha\beta}$ depend on
the types $\alpha$ and $\beta$ of beads $i$ and $j$, respectively.
The cutoff is set at $r^{c}_{\alpha\beta}=2^{1/6} \sigma$
to produce a purely repulsive interaction between beads.
For interactions between polymers of the same type
$\epsilon_{\alpha\alpha} = \epsilon$, with $\alpha=A$ or $B$.
The two types of polymers are made immiscible \cite{grest} by increasing the
repulsive energy between unlike beads
to $\epsilon_{AB} = \epsilon_{BA} = \epsilon(1 + \epsilon^*)$,
where $\epsilon^* = 1.2 $ or $3.2$. Both values of $\epsilon^*$ are sufficient 
to produce phase separation for the range of chain lengths $N$ considered
here, $N=16,$ $32,$ and $64$. 

The bounds of the simulation cell are periodic in the $x$ and $y$ directions,
with periods $L_x \cong 38.5 \sigma$ and $L_y \cong 33.4 \sigma$, 
respectively.
In the $z$ direction the cell is bounded by top and bottom walls.
Each contains $N_W=3200$ 
atoms tied to the sites of a $(111)$ plane of an fcc 
lattice by harmonic springs of stiffness
$\kappa = 1320\epsilon/\sigma^2$.
The walls are separated by $L_z\cong 47 \sigma$.
They confine $49152$ polymer beads, yielding a bead density of
$\rho= 0.8\sigma^{-3}$ in regions far away from interfaces.
Density oscillations are induced within a few $\sigma$ of the walls
\cite{khare}.
However, the solid/polymer interfaces are only introduced to produce
shear and are not of direct interest here. We thus restrict
our discussion to regions more than five $\sigma$ from the walls, where
wall-induced ordering is negligible.

A symmetric blend is created in the following way.
Polymers whose centers of mass are located in the bottom half of the
simulation box are labeled type $A$ and the remaining polymers are
labeled $B$.
The dividing plane is adjusted to ensure that there are equal
numbers of polymers of each type.
We also studied single phase systems with $N=10$ and $14$
at the same bead density.

Shear flow is induced by moving the top wall parallel to the interface
at a constant speed $v_{W}$ in the $x$ direction.
Due to the sharp wall/polymer interface and large polymer viscosity
there can be a substantial difference between the velocities of the wall
atoms and adjacent polymer beads
\cite{robbins,thompson,khare,jabbarzadeh,migler}.
We increased the wall/polymer interactions to limit
this interfacial ``slip''.
Atoms and beads interact through a Lennard-Jones potential with
$\sigma_{WA}=\sigma_{WB}=\sigma$.
For most simulations $\epsilon_{W\alpha} = \sqrt{1.7}\epsilon$
and the cutoff is extended to $r^{c}_{W\alpha}=\sqrt{1.5 }\sigma$
to include part of the attractive region in the Lennard-Jones potential.
At the highest shear rates studied, the wall/polymer interactions are increased
by choosing $\epsilon_{WA} = \epsilon_{WB} =\sqrt{4.7}\epsilon$, 
and $r_{c}=\sqrt{2.5 }\sigma$.

The equations of motion are integrated using  a 
fifth-order predictor-corrector method \cite{allen},
with a time step $\delta t = 0.0075 \tau$,
where $\tau = \sigma \sqrt{m/\epsilon}$ is the basic unit of time, 
and $m$ is the mass of a monomer.
A constant temperature of $k_BT=1.1\epsilon$
is maintained with a Langevin thermostat \cite{grest1}.
To ensure that this thermostat does not bias the shear profile, the
Gaussian white noise and damping terms are only added to the equations of
motion for the velocity components normal to the mean flow ($y$ and $z$)
\cite{robbins,stevens}.
There is a characteristic time required for energy to flow from
velocity fluctuations along the flow direction to the thermostatted
components.
When the inverse shear rate becomes comparable to this time, the
kinetic energy along the flow direction is no longer thermostatted
effectively.
This limits the maximum shear rate in our simulations to about
0.2$\tau^{-1}$.
However the correct ensemble for higher shear rates is a matter of
continuing debate, and such strongly nonequilibrium states are not
accessible to experiments \cite{stevens,evans}.

The local shear rate, $\dot{\gamma}$, of the fluid is easily calculated by
taking the local rate of change in the 
$x$ component of velocity, $v_x$ as a function of $z$, {\it i.e.} 
$\dot{\gamma} =  \partial v_x/ \partial z$.
This is done by taking slices parallel to the $x-y$ 
plane, of width $0.095 \sigma$,
and averaging the velocity of the monomers within these slices.
The viscosity within a slice is found from
\begin{equation}
\eta=\frac{P_{xz}}{\dot{\gamma}},
\end{equation}
where the shear stress $P_{xz}$ is constant throughout the system
in steady state.
Values of shear rate and viscosity presented below
are averaged over ten slices. 

\begin{figure}[hbt]
\epsfxsize=3.0in 
\centerline{\epsfbox{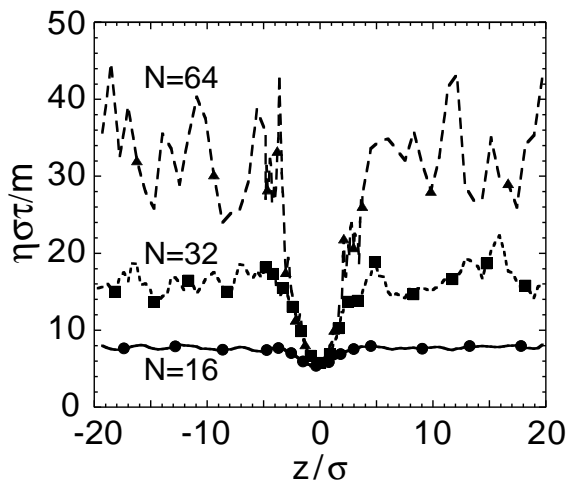}}
\caption{Newtonian viscosity as a function of height $z$
for $\epsilon^*=3.2$. The bulk viscosity scales linearly with chain length for
$N=16$ through $64$, while the interfacial viscosity $\eta_I$ is independent
of chain length over this range.
}
\label{vischeight}
\end{figure}

The wall velocity is varied from 
$v_w = 0.1 \sigma/\tau$ to $8.0 \sigma/\tau$. 
As shown in Table \ref{shear-all}, this leads to a variation by
almost two orders of magnitude in the bulk shear rate 
${\dot{\gamma}}_B$ evaluated far from any interface.
Note that this shear rate is the same in both fluids since we consider
a symmetric melt.
Statistical fluctuations drop with the total distance the wall moves
and increase with increasing chain lengths (e.g. Fig. \ref{vischeight}).
The simulations at lower shear rate
were sheared for more than a million time steps after equilibration.
Half this interval was used for higher shear rates.

\section{Results}\label{sec:result}
\subsection{Interface Width}
\begin{figure}[hbt]
\epsfxsize=3.0in 
\centerline{\epsfbox{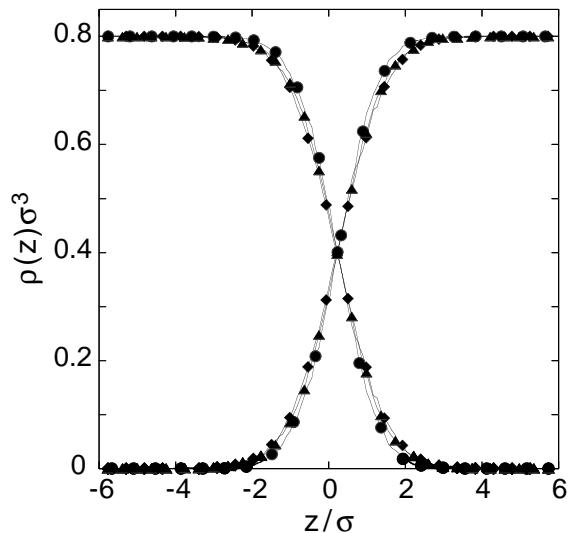}}
\caption{Densities of type $A$ (decreasing curves) and $B$ (increasing curves)
beads showing that the interface width is insensitive to chain length
and shear rate.
The immiscibility parameter is $\epsilon^*=3.2$ and
circles are for $N=64$, $v_W=0.1$; 
diamonds are for $N=64$, $v_W=5.0$;
and triangles are for $N=16$, $v_W=5.0$.
Corresponding values of ${\dot{\gamma}}_B$ are given in
Table \ref{shear-all}.
}
\label{dens}
\end{figure}

In Fig.\ref{dens} we show the densities of the two types of beads in the
region near the interface for $\epsilon^* =3.2$ and $N=16$ and 64.
For this strongly immiscible case,
the densities change from their bulk values to zero over
a few $\sigma$.
Moreover, these density profiles are nearly unaffected when $N$ is changed
by a factor of $4$ and the bulk shear rate is changed by a factor of 100.

Previous studies of equilibrium interfaces have related the interface
width to the radius of gyration of polymer segments that enter the
interfacial region \cite{flory,degennes,goveas,helftag,fur}.
The length of these segments, $s^*$, is determined by the degree of
immiscibility and becomes independent of $N$ in the large $N$ limit.
Earlier studies of the equilibrium properties of the model considered
here\cite{barsky} are consistent with these predictions.
For example, for $\epsilon^*=3.2$ the density profiles are independent of $N$
for $N \geq 16$.
As illustrated in Fig. \ref{dens}, this behavior extends to the highest
shear rates studied here.
The same insensitivity to shear rate is found for the less immiscible
case of $\epsilon^*=1.2$ where the interface is slightly wider.

\subsection{Viscosity}
The bulk viscosity of polymers whose length is shorter than the entanglement
length can be described with Rouse theory \cite{doi}.
The limiting Newtonian viscosity at low shear rates scales linearly
with chain length: $\eta_B = \zeta b^2 \rho N $, where $\zeta$ is the
monomeric friction coefficient and $b$ is the statistical 
segment length.
DeGennes and coworkers \cite{degennes}
argued that the viscosity in the interfacial region, $\eta_I$,
should be determined by the effective length $s^*$ rather than
$N$ because $s^*$ is the length of segments that must relax 
during shear flow.
Their conclusion is supported by the analysis of 
Goveas and Fredrickson\cite{goveas} and by our earlier simulations
in the Newtonian limit \cite{barsky}.

The dependence of the bulk and interfacial viscosities on chain length
in the Newtonian limit is illustrated in Figure \ref{vischeight}.
Here $\eta$ is plotted against height $z$ for $\epsilon^*=3.2$
over the range of chain lengths where 
the interface width, and thus $s^*$, is constant.
Away from the interface $\eta$ approaches the bulk viscosity,
which rises linearly with $N$ for $N$ up to 64.
The slope of this rise
determines $\zeta=0.38\pm 0.03 m/\tau$, since $b=1.28\sigma$
is known from static properties \cite{barsky,grest1}
and $\rho=0.8\sigma^{-3}$.

In the interfacial region the viscosity is smaller than the bulk
value and is independent of chain length.
This is consistent with the picture that the constant value of
$s^*$ determines $\eta_I$.
Equating $s^*$ to the chain length that would give a bulk viscosity
equal to $\eta_I$ yields $s^* \approx 10$.
A similar analysis for $\epsilon^*=1.2$ yields $s^* \approx 14$,
and in both cases the radius of gyration corresponding to $s^*$
is 1.6 times the interface width \cite{barsky}.
Thus the Newtonian response is consistent with the picture
advanced by deGennes and coworkers \cite{degennes}.

\begin{figure}[hbt]
\epsfxsize=3.0in 
\centerline{\epsfbox{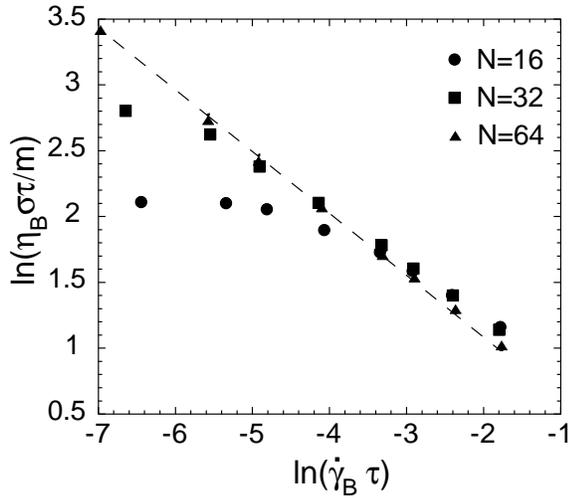}}
\caption{Bulk viscosity $\eta_B$ as a function of shear rate
${\dot{\gamma}}_B$, for polymers of length $N=16,32$ and 64.
As $N$ increases, shear thinning is observed at lower values
of ${\dot{\gamma}}_B$.
The dashed line shows a power law fit $\eta_B \sim {\dot{\gamma}}^{-0.47}$. }
\label{viscbulk}
\end{figure}

The shear thinning behavior of the bulk viscosity with
increasing shear rate is shown in Fig. \ref{viscbulk}.
As is well known, the onset of shear thinning occurs when the
system is sheared more rapidly than it can relax.
Since the time to relax increases with chain length, the
onset of shear thinning moves to lower shear rates with increasing
$N$ \cite{xu}.
As in previous work \cite{xu}, the decrease in viscosity at high
shear rates can be fit to a power law $\eta \sim {\dot{\gamma}}^{-\alpha}$,
with $\alpha$ near 1/2.
The dashed line in Fig. \ref{viscbulk} shows a fit to
$N=64$ data with $\alpha = 0.47$.
The low shear rate viscosity increases with chain length,
and the high shear rate behavior is nearly independent of $N$.

\begin{figure}[hbt]
\epsfxsize=3.0in 
\centerline{\epsfbox{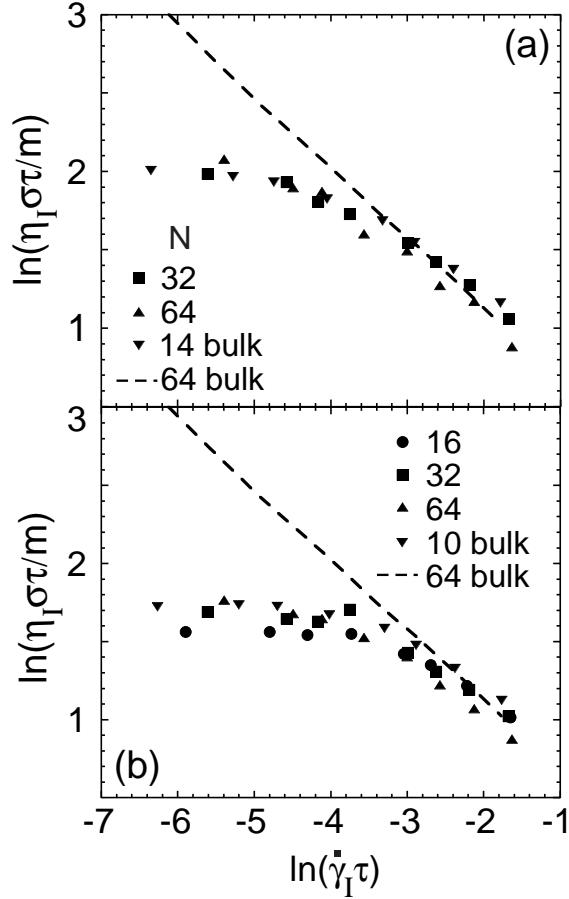}}
\caption{
The interfacial viscosity $\eta_I$ as a function of 
interfacial shear rate ${\dot{\gamma}}_I$ for (a) $\epsilon^*=1.2$ and
$N=32$ and 64, and (b) $\epsilon^*=3.2$ and $N=16$, 32 and 64.
The bulk viscosities of polymers with length equal to the values of $s^*$
determined from the Newtonian response are shown for comparison.
The dashed lines show the bulk viscosity of chains of length $64$.
Interfacial shear rates are evaluated over an interval of width one $\sigma$
in the center of the interface.
}
\label{ifvisc}
\end{figure}

The variation of the interfacial viscosity with the interfacial
shear rate is shown in Fig. \ref{ifvisc}
for different $N$ and $\epsilon^*$.
In all cases, $\eta_I$ begins to shear thin at much higher shear rates
than $\eta_B$.
Data is only presented for values of $N$ where the Newtonian response
and interface width are independent of chain length, so that $s^*$
should also be constant.
One sees from the figure that the shear thinning behavior is also 
independent of $N$ in this limit.
Moreover, the entire shear thinning curve for all $N$ at
each value of $\epsilon^*$
is consistent with the bulk shear thinning behavior of
chains with length equal to
the values of $s^*$ inferred from the corresponding Newtonian response.
This is strong evidence that a single segment length determines
the non-Newtonian shear thinning behavior as well as the static
interface profile and Newtonian response.

\subsection{Polymer Conformations}
\begin{figure}[hbt]
\epsfxsize=3.0in
\centerline{\epsfbox{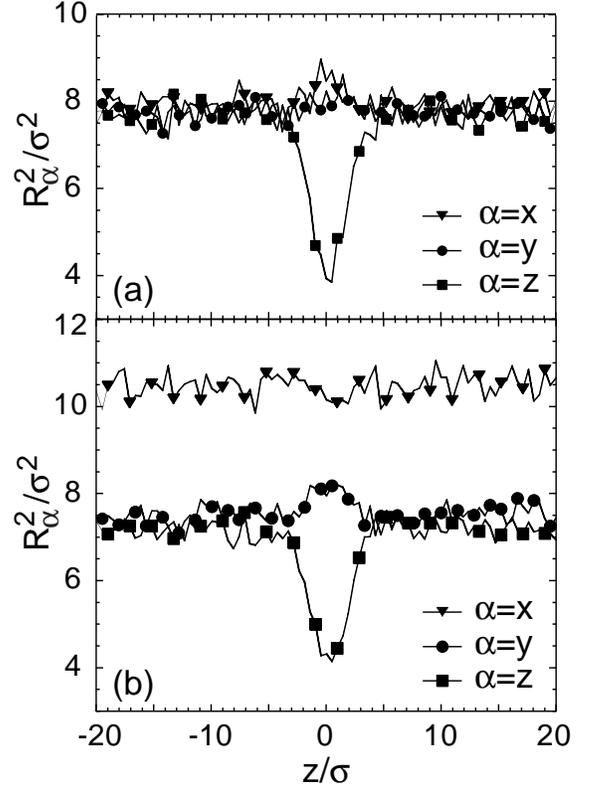}} 
\caption{The mean square components of the end-to-end distance along
$\alpha= x$, $y$, and $z$ as a function of the center of mass
height $z$ for $N=16$ polymers with $\epsilon^*=3.2$ at
(a) $v_W=0.1\sigma/\tau$ and (b) $v_W=0.5\sigma/\tau$.
Although the viscosities at both shear rates are nearly indistinguishable
from the limiting Newtonian viscosity,
there is a  clear stretching of polymers along the flow direction in (b).
}
\label{n=16conformation}
\end{figure}

We now describe the changes in chain conformation that
are associated with the shear thinning of the bulk and interface.
Fig. \ref{n=16conformation} shows the mean square of the
three components of the end-to-end vector $\vec{R}$ as a function of
the height of the polymer's center of mass.
The two panels show results for two low shear rates
with $N=16$ and $\epsilon^*=3.2$.
The lowest shear rate (panel (a)) is well into the Newtonian regime, and
the polymers have time to relax to equilibrium conformations.
Far from the interface the polymers follow isotropic
random walks with $R^2_x=R^2_y=R^2_z \approx 8\sigma^2$.
Near the interface $R^2_z$ is strongly suppressed because immiscibility
eliminates paths that take polymers too far into the other phase.
The values of $R^2_x$ and $R^2_y$ are nearly unaffected by this constraint,
but show a small rise that is comparable to our statistical noise.
The value of $R^2_z$ does not attain its bulk level until the center
of mass is far enough from the interface that chains are unlikely
to reach it.
This distance is comparable to the bulk value of $\sqrt{R^2_z}$,
which is larger than the range over which the density changes
(Fig. \ref{dens}).

Panel (b) of Fig. \ref{n=16conformation} shows the mean square
components of $\vec{R}$
at the third lowest shear rate in Fig. \ref{viscbulk}.
Surprisingly,
although the viscosity is nearly indistinguishable from the low
shear rate limit,
there is a substantial change in conformation.
The component along the flow direction, $R^2_x$,
increases by about 25\% throughout the system.
The bulk value of $R^2_z$ drops by about 10\%, while the interfacial
value is nearly unchanged.
This is consistent with the observation that the interface shear
thins at higher shear rates than the bulk.

\begin{figure}[hbt]
\epsfxsize=3.0in
\centerline{\epsfbox{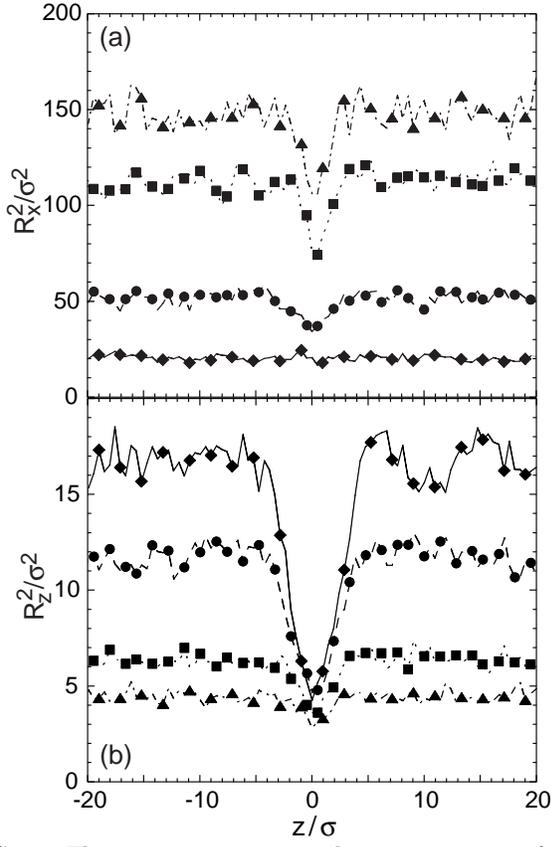}}
\caption{The mean square $x$ and $z$ components of
the end-to-end vector as a function of center of mass
height for $N=32$ polymers at wall velocities
$v_W=0.1\sigma/\tau$ (diamonds), $0.5 \sigma/\tau$ (circles),
$2\sigma/\tau$ (squares) and $5\sigma/\tau$ (triangles).
Corresponding
bulk shear rates are given in Table \ref{shear-all}.
}
\label{r2n32xyz}
\end{figure}

Conformation changes at much higher shear rates are illustrated in
Fig. \ref{r2n32xyz}.
Here the mean square $x$ and $z$ components of the 
end-to-end vector are plotted as a function of
center of mass height for $N=32$, $\epsilon^*=3.2$,
and four values of $v_W$.
As expected, chains in the bulk
regions are stretched and aligned along the flow direction.
At the lowest shear rate,
the bulk value of $R^2_x$ is about 20\% above its equilibrium value.
At the highest shear rate, the length of the polymer along the flow
direction has grown to about 40\% of the fully extended length.
The components orthogonal to the flow decrease.
The drop in $R^2_z$ is much larger than that in $R^2_y$,
which drops from about 17 to 10$\sigma^2$ over the studied 
range of $\dot{\gamma}$.
Overall there is a net increase in $R^2$ from about 50 to 180$\sigma^2$,
because the increase 
in $R^2_x$ is much larger than the decrease in the other two components.

Chains in the interfacial region also stretch along the flow
direction and contract in the orthogonal directions.
However, as with the viscosity, the shear rate dependence is shifted
to higher $\dot{\gamma}$ at the interface.
The value of $R_x^2$ is initially slightly higher at the interface
and increases more slowly with $\dot{\gamma}$, leading to a pronounced
dip at the highest shear rates.
In contrast,
the dip in $R_z^2$ at the interface decreases with increasing shear
rate because the bulk value drops more rapidly with $\dot{\gamma}$.

\begin{figure}[hbt]
\epsfxsize=3.0in
\centerline{\epsfbox{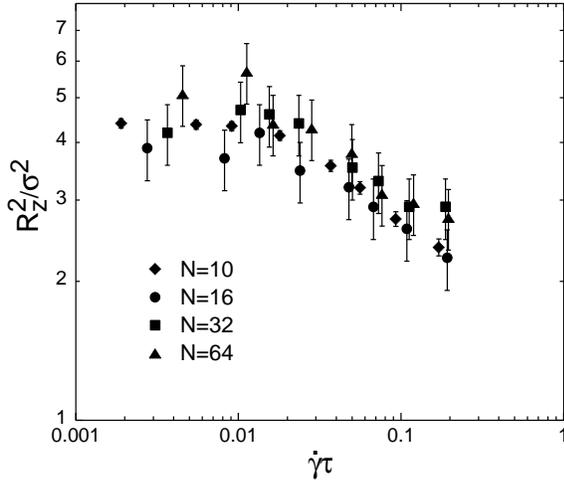}}
\caption{The mean square $z$ component of the end-to-end vector, $R^2_z$,
for polymers centered at the interface with $N=16$, 32 and 64.
Bulk values for chains with $N=10$ are shown for comparison. 
All curves begin to decrease at the same shear rate, and this 
shear rate coincides with the drop in interfacial viscosity.
}
\label{r2zall}
\end{figure}

Fig. \ref{r2zall} provides a more detailed picture of
the variation in the conformation of
chains near the interface with increasing shear rate.
Interfacial values of $R^2_z$ are plotted against shear rate for
$N=16,$ 32, and 64 at $\epsilon^*=3.2$.
Bulk values for chains with length equal to the inferred value
of $s^*=10$ are shown for comparison.
All curves are flat at low shear rates, and begin to decrease at
the point where the interfacial viscosity begins to fall in
Fig. \ref{ifvisc}.
As with the data for $\eta_I$,
the interfacial conformations for all chain lengths collapse
on to a universal curve that coincides with the bulk curve for
chains of length $s^*$ within our errorbars.
This provides further evidence that a single time scale corresponding
to $s^*$ controls the behavior of the interface.
Note that the uncertainties are larger for interfacial values
due to the relatively
low number of polymers whose center of mass lies at the interface.
As discussed previously\cite{barsky,helfand}, there is a preponderance
of chain ends at the interface.
This reduces the number of chain centers at the interface
by a factor of three for the parameters considered here.

The idea that relaxation times should scale with the length of
a subset of the chain rather than its entire length is related
to other recent work\cite{stuff}.
These papers have examined changes in the conformation
of individual chains pulled through gels or other polymers.
Segments at the free end of the chain are assumed to
relax with a time scale that is characteristic of bulk
polymers with the same length as the segment.
Since the relaxation time grows with the length of the segment,
the polymer is less relaxed and more stretched
as one moves away from the free end toward the pulling end.

\begin{figure}[hbt]
\epsfxsize=3.0in
\centerline{\epsfbox{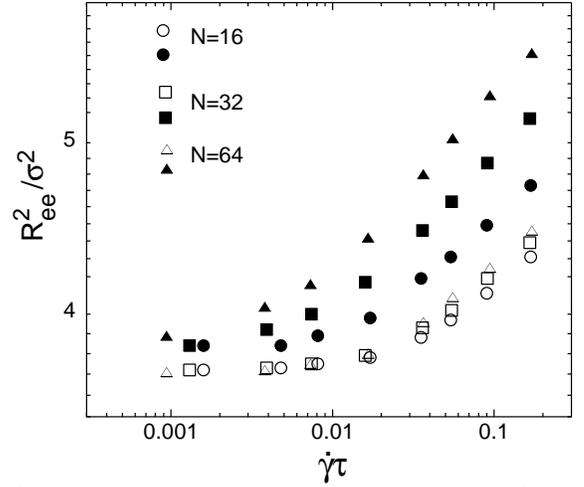}} 
\caption{The mean square end-to-end distance of 4 bead segments at the middle 
(filled symbols) and ends (open symbols) of polymers with the indicated
$N$ as a function of shear rate.
The ends show the same behavior for all $N$.
The middles begin to stretch at a lower shear rate that decreases
with increasing $N$.
}
\label{subsect}
\end{figure}

A similar phenomenon should occur within individual chains in a sheared
polymer melt.
Segments at the end of the chain will be more relaxed than the center
leading to a dumbbell configuration that narrows in the center.
In addition, the ends should begin to stretch at higher shear rates
than the center because their characteristic
relaxation time is shorter.
Fig. \ref{subsect} shows the end-to-end distance of segments of length $4$
as a function of shear rate.
Results for the two end-segments are averaged and compared to a segment in the
middle of the chain.
Note that the ends show the same behavior for all $N$ indicating that
the characteristic relaxation time for their conformations is only
a function of segment length.
In contrast, the middle segment becomes more stretched as $N$ increases
and the characteristic rate at which it can no longer relax decreases
with increasing $N$.
These effects are reminiscent of a nematic coupling effect seen
recently in polymers,
where the orientation of middle segments also relaxed more slowly 
with increasing $N$, while the rapid relaxation of ends was independent
of $N$ \cite{mem}.

\section{Summary and Discussion}\label{sec:dis}

In this paper we have examined shear thinning and conformational
changes in the bulk and interfacial regions of a phase-separated
binary blend of Rouse chains.
The bulk behavior is consistent with previous studies.
The most surprising result is that pronounced elongation of the chains
is observed before there is a noticeable change in viscosity.
The onset of significant changes in $\eta_B$ seems more closely
connected to the contraction along the gradient direction,
which becomes appreciable
at higher shear rates than the elongation along the flow direction.

The changes in interfacial viscosity are consistent with the prediction
\cite{degennes,goveas} that the interfacial viscosity is determined entirely
by the characteristic length $s^*$ of loops that cross
the interface between immiscible polymers.
This length is determined by the degree of immiscibility and independent
of $N$ for the polymers studied here.
The interfacial viscosity shows these same trends.
For each degree of immiscibility, plots of interfacial
viscosity against interfacial shear rate for different $N$
collapse onto a single curve.
The entire curve coincides with the bulk shear thinning behavior
of polymers of length $10$ for $\epsilon^*=3.2$ and $14$ for
$\epsilon^*=1.2$.
Thus a single effective chain length describes both the Newtonian value
of $\eta_I$ and its shear thinning behavior.
Our previous work showed that the same effective chain length described
the width of equilibrium interfaces.\cite{barsky}

The suppression of the viscosity near the interface (Fig. \ref{vischeight})
leads to an
effective slip boundary condition \cite{degennes,goveas,barsky}.
The amount of slip is most pronounced in the Newtonian limit where
the difference between $\eta_I$ and $\eta_B$ is largest.
As the bulk viscosity begins to shear thin, the amount of slip
decreases.
Once the shear rate is high enough to produce significant shear thinning
at the interface, the bulk and interfacial viscosities converge
(Fig. \ref{ifvisc}).
In this limit the amount of slip is negligible.

Changes in the conformation of polymers near the interface were also
studied.
In equilibrium,
the interface suppresses the component of the end-to-end vector
perpendicular to the interface, but has little effect on the in-plane
components.
The changes in interfacial viscosity are most closely correlated with
changes in the perpendicular component.
Like $\eta_I$,
plots of $R_z^2$ against shear rate for different
chain lengths collapse onto a common curve that is consistent
with that for bulk chains of length $s^*$ (Fig. \ref{r2zall}).
While the elongation increases more slowly near the interface than in the bulk,
it begins at much lower shear rates than the change in interfacial
viscosity (i.e. Fig. \ref{n=16conformation}) and depends on chain length.
Thus our results for both the bulk and interface indicate
that polymer contraction perpendicular to the interface
is the most important structural change associated with shear thinning.

We also explored changes in the conformation of segments of 4 beads
within longer chains as a function of shear rate.
Segments in the center of the chains began to elongate at the same shear
rate where the bulk viscosity for the given chain length showed shear thinning.
However, the conformation of the ends remained unchanged until much higher
shear rates.
Moreover, the conformation of the ends was nearly independent of the total
chain length and their shear thinning behavior was comparable to that for
bulk chains with length 4.
Recent studies of much longer chains have used the concept of a position
dependent relaxation time to determine the conformation of a single
chain in a solvent or gel \cite{stuff}.
Our study shows that the same concept applies to polymer melts.

\section{Acknowledgements} 
Support from the Semiconductor Research Corporation through AMD Custom
Funding and from National Science Foundation Grant No. DMR 0083286 is
gratefully acknowledged.

\begin{table}
\caption{
The shear rate in bulk regions of each polymer ${\dot{\gamma}}_B$
at the indicated values of chain length and wall velocity $v_W$.
Uncertainties are less than $5 \%$.
The proportionately larger shear rates at $ v_W=8 \sigma/\tau$
are due to an increase in the wall-coupling parameters,
as discussed in Sec. \ref{sec:method}.}
\begin{tabular}{lccccc}
$v_W$ & & & $\dot{\gamma}_B$ &\\
 & $N=10$ & $N=14$ & $N=16$ & $N=32$ & $N=64$ \\
$0.1$   & $ .0019$ & $ .00175$  &  .00159  & $0.0013$  & $0.00094$  \\
$0.3$   & $ .0055$ & $  .00512$    & .00479  & $0.0039$  &  $0.0038$ \\
$0.5$   & $ .0091$ & $ .00872$     & $0.0081$  & $0.0074$  &  $0.0073$ \\
$1.0$   & $ .018$ &  $ .00175$   & $0.0171$   & $0.0159$ &  $0.0166$  \\
$2.0$   & $ .037$ & $  .036$   & $0.0352$   & $0.0360$  & $0.0365$  \\
$3.0$   & $ .0563$ & $ .0549$    & $0.0538$   & $0.0545$ &   $0.0554$  \\
$5.0$   & $ .093$ & $ .0908$    & $0.0899$   & $0.0907$ &   $0.0943$  \\
$8.0$   &  $.171  $  & $.168  $ & $0.168$   & $0.166$ &   $0.171$
\end{tabular}
\label{shear-all}
\end{table}


\end{document}